\documentstyle[aps,preprint]{revtex}
\draft
\input epsf
\epsfverbosetrue
\begin{document}
\title{Combination of random-barrier and random-trap models}
\author{K. Mussawisade, T. Wichmann, and K.W. Kehr\\
Institut f\"ur Festk\"orperforschung, Forschungszentrum
J\"ulich GmbH,\\ 52425 J\"ulich, Germany}
\date{\today}
\maketitle
%
\begin{abstract}
The temperature dependence of the diffusion coefficient of particles is
studied on lattices with disorder. A model is investigated
with both trap and barrier disorder that was introduced
before by Limoge and Bocquet
(1990 Phys.\ Rev.\ Lett.\ {\bf 65} 60) to explain an
Arrhenian temperature-dependence of the diffusion coefficient
in amorphous substances.
We have used a generalized effective-medium approximation (EMA) by
introducing weighted transition rates as inferred from
an exact expression for the diffusion coefficient
in one-dimensional disordered chains.
Monte Carlo simulations were made to check the validity of the
approximations. Approximate Arrhenian behavior can be achieved in
finite temperature intervals in three- and higher-dimensional
lattices by adjusting the relative strengths of the barrier and
trap disorder. Exact Arrhenian behavior of the diffusion coefficient
can only be obtained in infinite dimensions.
\end{abstract}
%
\newpage
\section{Introduction}
Many amorphous substances exhibit linear behavior in an Arrhenius
plot of
the diffusion coefficient $D$, or the mobility $B$ of particles, where
$\ln (D)$, or $\ln (B)$, is presented
as a function of the inverse temperature $\beta=1/Tk_B$. The observation
of an Arrhenian temperature-dependence
is not easily understandable from the theory of diffusion in disordered
crystals.
The commonly used models of diffusion of particles in lattices with
disordered transition rates predict
different behavior: The random-trap model predicts generally convex
(downward) curvature of $\ln (D)$ versus $\beta$,
independent of the lattice dimension $d$, while
the random-barrier model gives concave (upward)
curvature for three- and higher-dimensional lattices.
Limoge and Bocquet \cite{LB} suggested that the apparent Arrhenius
behavior of $\ln (D)$ versus $\beta$ might be due to a compensation of
the  effects of random barriers and random traps.
The aim of this paper is the examination of this appealing proposition
by analytical and numerical methods.

The analytical arguments given by Limoge and Bocquet \cite{LB} in
support of a compensation of the effects of random barriers and
of random traps
are only partially satisfactory. They employed continuous-time
random-walk theory
and used a decoupling approximation when performing the disorder average.
This procedure
gives correct results for the random-trap model. For the random-barrier
model, however, the ensuing
results are not correct, as a consequence of the neglect of important
backward correlations
in the transitions of the particles. 
The failure of the decoupling approximation for the random-barrier model in 
comparison to the Monte Carlo results was already discused in \cite{LB}.

Recently a new exact result for the
diffusion coefficient of
particles in one-dimensional disordered chains
has been found \cite{wich,diet,kut}. The expression contains transition
rates that are weighted by
the equilibrium occupancies of the sites; this is relevant for
random site energies, i.\ e., random traps.
The insight obtained from this result can serve as a basis for
approximate treatments 
of models with site and barrier disorder
in higher dimensions, as will be shown here
in Form of an effective-medium approximation (EMA).
Very recently Limoge and Bocquet \cite{LBtobep} tried to take the inherent
backward correlations of the random-barrier model into account. The 
differences between their and our results will be discussed below.

In the remainder of the Introduction we present qualitative arguments
for the different curvatures in
the simple models. The downward curvature of $\ln (D)$ versus $\beta$ for the
random-trap model is
easily understood:  The deepest trap sites with the lowest energies
dominate the behavior at the
lowest temperatures. The convex curvature can be deduced formally from
the exact expression for the disorder-averaged diffusion coefficient
$D$ (see below), which is valid in all dimensions $d$.
The argument for the random-barrier model is more
complicated. At the
lowest temperatures, where the ratio of the width of the  barrier energy
distribution and the thermal energy is large, the
diffusion coefficient is determined by the highest barrier along a
critical path of bond percolation \cite{AHL}. The path
may be constructed by
selecting successively bonds with the lowest barriers possible until the
``infinite'' cluster in the lattice appears. If the temperature
is raised,
additional paths contribute to the diffusivity. These paths comprise also
higher barriers, hence
the apparent activation energy is increased.

In the following section we describe the derivation of the exact result
for $D$ in disordered linear chains from an expression for
the mean first-passage time. In Sec.\ 3 we give an analytical
treatment of the
combination of independent random barriers and random traps. Sec.\ 4
contains the EMA for the combined model in $d \ge 3$.
In Sec.\ 5 we present our conclusions from the results.
\section{Exact expression for diffusion coefficient in $d = 1$}
Recently an exact expression for the asymptotic diffusion coefficient of
single particles on
linear chains with disordered transition rates has been derived
\cite{wich,diet,kut}. In
\cite{wich} a first-passage time method was used while in \cite{diet,kut}
the mobility
of particles on chains with periodic boundary conditions was derived
from the linear response
to a driving force. We follow reference \cite{wich} for the derivation.
The basis is an exact expression for the mean first-passage time of a
particle from
site 0 to site $N$ on a segment of a disordered chain \cite{murthy}.
Site 0 is regarded as reflecting,
while site $N$ is an absorbing site. The expression reads
\begin{equation}
\label{gt0n}
\bar{t}_{0N} =
\sum_{k=0}^{N-1}\frac{1}{\Gamma_{k, k+1}}+
\sum_{k=0}^{N-2}\frac{1}{\Gamma_{k, k+1}}
\sum_{i=k+1}^{N-1}\prod_{j=k+1}^{i}
\frac{\Gamma_{j, j-1}}{\Gamma_{j, j+1}}.
\end{equation}
Here $\Gamma_{i,i+1}$ is the transition rate from site $i$ to site $i+1$;
the transitions being restricted to nearest neighbors. Note
that (\ref{gt0n}) gives
$\bar{t}_{0N}$ for a particular (quenched)
realization of the disordered segment.

We now invoke the condition of detailed balance between two neighbor
sites,
\begin{equation}
\label{detb}
\rho_i\Gamma_{ij} = \rho_j\Gamma_{ji}.
\end{equation}
The thermal occupation factors $\rho_i$ are defined by
\begin{equation}
\rho_i = \frac{\exp (-\beta E_i)}{\{\exp (-\beta E_i)\}}  \label{occ}
\end{equation}
where $E_i$ is the energy of site $i$, counted negative
from a common origin.
The curly brackets in  (\ref{occ}) designate the average over the
disordered site energies, for finite segments
\begin{equation}
\{\exp (-\beta E_i)\} = \frac{1}{N} \sum^{N}_{i=1} \exp(-\beta E_i).
\end{equation}
The occupation factors are proportional to the thermal equilibrium
occupation probabilities of
the sites; due to the normalization used they can be larger or smaller
than one.

The condition of detailed balance holds in equilibrium, and the
occupation factors $\rho_i$ exist
for finite segments when all nearest-neighbor
transition rates $\Gamma_{ij} \neq 0$. We have to require that
unique occupation factors also exist in the limit $N \to \infty$. This
requirement
excludes certain interesting models, for instance the Sinai model
\cite{sinai}, from the further derivations. However, it is
fulfilled for the models considered in the next section.

When the condition of detailed balance (\ref{detb}) is introduced into
expression  (\ref{gt0n})
a considerable simplification is achieved:
\begin{equation}
  \bar{t}_{0N} =
  \sum_{k=0}^{N-1}\frac{1}{\Gamma_{k, k+1}}+
  \sum_{k=0}^{N-2}  \rho_k
  \sum_{i=k+1}^{N-1}
  \frac{1}{\rho_i\Gamma_{i, i+1}}.
\end{equation}
Taking the disorder average of the equation, we find as the leading term
in the limit of long segments,
$N >> 1$,
\begin{equation}
\label{gt0nlimit}
\left\{ \bar{t}_{0N}\right\} =
\frac{1}{2}N^2\left\{\frac{1}{\rho_i \Gamma_{ij}}
\right\}.
\end{equation}
This equation can be interpreted as the inverted relation between time
and mean-squared
displacements of random walks on disordered lattices, $t =
(2D)^{-1}\{X^2\}$.
We hence deduce the following asymptotic diffusion coefficient from
(\ref{gt0nlimit})
\begin{equation}
\label{dex1}
D = \left\{ \frac{1}{\rho_i \Gamma_{ij}} \right\}^{-1}.
\end{equation}
Under the assumption of the existence of unique occupation factors one
can also derive, for large $N$, using again detailed balance
\begin{equation}
\{ \bar{t}^2_{0N}\} - \{ \bar{t}_{0N}\}^2 \propto N^3.
\end{equation}
Since $\{\bar{t}_{0N}\}^2 \sim N^4$, the relative dispersion of the
disorder average of the mean
first-passage times vanishes as $N^{-1/2}$ for long segments. One can say
that $\{\bar{t}_{0N}\}$
becomes ``sharp'' in the asymptotic limit of large $N$.
In this way the use of the inverted  relation  (\ref{gt0nlimit})
to deduce the diffusion coefficient can be justified.

The physical significance of the result  (\ref{dex1}) is that the
diffusion coefficient follows from thermally
weighted transition rates. A numerical
verification of (\ref{dex1}) for
the Miller-Abrahams model \cite{MA} in $d=1$ has been
given in reference \cite{conf}.
\section{Combination of Random Barriers and Random Traps: Analytical
Results}
\subsection{Model and exact result in $d = 1$}
\label{kresult1d}
We now investigate a model for hopping diffusion of particles in
lattices with combinations of random barriers and random
traps. The random-barrier (RB) model is defined with symmetric
transition rates $\Gamma_{ij}$ between neighbor
sites, $\Gamma_{ij} = \Gamma_{ji}$. An Arrhenius form for the transition
rates is assumed,
\begin{equation}
\Gamma_{ij} = \Gamma_0 \exp (-\beta E_{ij}).
\end{equation}
The energetic barrier $E_{ij}$ between sites $i$ and $j$ is a random
variable and it is taken from a common distribution
$\nu_B(E)$. To avoid problems with negative barriers we restrict the
range of the barrier energies to $E \ge 0$. The random-trap
(RT) model is defined by rates $\Gamma_i$ that originate from the sites
$i$  and are independent from the final sites. Also
here the Arrhenius form is assumed,
\begin{equation}
\Gamma_i = \Gamma_0 \exp(\beta E_i)
\end{equation}
where $E_i$ is the energy of site $i$. The site energies will be counted
negative and the range be restricted to $E \le 0$.
Again the individual site energies are selected from a common
distribution $\nu_T(E)$.

We introduce a combination of the
RT and RB models by specifying the transition rates between two
neighbor sites as
\begin{equation}
\Gamma_{ij} = \Gamma_0 \exp [-\beta (E_{ij} - E_i)]
\end{equation}
The energy $E_{ij}$ with two site indices refers to the barrier
$i \rightarrow j$ while the energy $E_i$ with one index refers
to the site $i$. (To distinguish between barrier and site energies,
we sometimes keep dummy site indices).
A pictorial representation of the model is given in figure 1.
Note that the model is different from the Miller-Abrahams
model \cite{MA} where the site energy of the terminal site $j$ appears
explicitly.

The occupation factors $\rho_i$ are required in the weighted transition
rates that appear in  (\ref{dex1}); they are
given in  (\ref{occ}). The weighted transition rates are then
\begin{equation}
\rho_i\Gamma_{ij} = \frac{\Gamma_0 \exp (-\beta E_{ij})}{\{\exp (-\beta
E_i)\}}.  \label{trw}
\end{equation}
The numerator contains only barrier energies while the site energies
only appear in the denominator, which can be evaluated directly.
An immediate consequence of (\ref{trw}) is the
expression for the diffusion coefficient of the RT model,
\begin{equation}
D_{RT} = \Gamma_0 \{ \exp(-\beta E_i)\}^{-1}   \label{rtd}
\end{equation}
when barrier disorder is absent. Equation (\ref{rtd}) follows from
(\ref{dex1}) in $d = 1$; it is valid in all
dimensions \cite{HK}. Another consequence of Eq.\ (\ref{rtd})
is that the derivative of $\ln (D_{RT})$ with respect to the
inverse temperature $\beta $ is given by the mean thermal energy.
Since the mean thermal energy can only decrease with decreasing
temperature, the slope of $\ln [ D_{RT}(\beta )]$ is decreasing with
increasing $\beta $.

An exact result is obtained from (\ref{dex1}) for the diffusion
coefficient of the combined RB and RT model in $d = 1$,
\begin{equation}
D_{comb} = \Gamma_0 \{ \exp (-\beta E_i) \}^{-1} \{ \exp
(\beta E_{ij})\}^{-1}.
 \label{comb1}
\end{equation}
Equation (\ref{comb1}) can be cast into another form,
\begin{equation}
D_{comb} = \frac{1}{\Gamma_0} D_{RT} D_{RB}.  \label{comb2}
\end{equation}
The diffusion coefficient $D_{RT}$ exhibits generally downward curvature
in an Arrhenius plot. Since the average
that determines the diffusion coefficient $D_{RB}$ in $d = 1$ has the
same form
as the one determining $D_{RT}$, downward curvature is also present
in the Arrhenius plot of this coefficient. Also the diffusion coefficient
of the combined model has then downward
curvature in $d = 1$. In other words, a compensation of the effects of
random barriers and random traps is never possible in $d = 1$.

\subsection{Results in higher dimensions}

The diffusion coefficient of particles in the random-barrier model in
simple-square lattices is exactly known for
symmetric barrier-energy distributions $\nu_B(E)$ \cite{berna}. For
symmetric
energy distributions the diffusion coefficient is
\begin{equation}
  D_{RB} = \Gamma_0 \exp(-\beta \bar{E})
  \label{gdrb2dsym}
\end{equation}
where $\bar{E}$ is the medium value of the energy distribution.
Hence $D_{RB}$ does not show curvature in an Arrhenius plot.
Consequently, the diffusion coefficient of the combined
model will exhibit downward curvature in simple-square lattices.

Approximations are necessary to derive the asymptotic diffusion
coefficients in higher-dimensional disordered lattices.
It is very plausible that the weighted transition rates (\ref{trw})
should also be used in approximate treatments.
Site-energy and barrier disorder can be treated independently when they
are  uncorrelated. Under this assumption, the numerator of
Eq.\ (\ref{trw}) only contains uncorrelated barrier energies.
The EMA gives reasonably accurate results for the RB model
when the disorder is not very strong. Hence we will
use this approximation to deal with
the barrier disorder in $d > 1$. In the
EMA, an effective transition rate $\Gamma_{eff}$ is determined
from a self-consistency condition \cite{berna,kirk},
\begin{equation}
\left\{ \frac{\Gamma_{eff} - \Gamma}{\frac{z-2}{2} \Gamma_{eff} + \Gamma}
\right\} = 0. \label{self}
\label{gselfconsis}
\end{equation}
The transition rate $\Gamma$ is a random variable, here it is taken
according to Eq.\ (\ref{trw}). The rate $\Gamma$ is
symmetric as a consequence of detailed balance; this symmetry is
required for the application of Eq.\ (\ref{self}).
The curly brackets indicate the disorder average which extends over
the barrier disorder.
The diffusion coefficient is identical to $\Gamma_{eff}$ and the
lattice constant in the hypercubic lattices that we study
is set unity.

Since the exact result (\ref{rtd}) for $D_{RT}$ is
contained as a factor in the weighted transition rates, we
have in  EMA
\begin{equation}
D_{comb}^{EMA} = \frac{1}{\Gamma_0} D_{RT} D^{EMA}_{RB}.   \label{comb3}
\end{equation}
The EMA becomes exact in the limit of coordination number
$z \to \infty$, i.\ e., in infinite-dimensional disordered
lattices. We obtain in this limit
\begin{equation}
D_{comb} = \Gamma_0 \{ \exp (-\beta E_i)\}^{-1} \{ \exp(-\beta E_{ij})\}.
\end{equation}
The second average is proportional to the average over the rates
corresponding to the random barriers. For the RB model and in
$d \to \infty$ we have $D_{RB} = \{ \Gamma_{RB}\}$, hence the product
form of $D_{comb}$ (\ref{comb2}) is also
valid in infinite dimensions. We conjecture that (\ref{comb2}) is
generally valid if the site and barrier
energies are uncorrelated.
Very recently Limoge and Bocquet incorporated the backward correlations of
the RB model and derived a self-consistency condition for the effective 
transition rate. However, their self-consistency condition is different from 
the standard form (\ref{gselfconsis}) of the EMA and does neither reproduce the
exact one-dimensional (section \ref{kresult1d}) nor the exact 
two-dimensional result (\ref{gdrb2dsym}).

\subsection{Possibility of complete compensation}

In this subsection we study the possibility for a complete compensation
of the curvatures resulting from the RT and RB models. For this purpose
we consider first the case of infinite dimensions, $ d \rightarrow
\infty$. To obtain complete compensation of the curvatures, a special
relationship between the site and barrier energy distributions must
exist. To determine this relation, we require that the following
equation represents simple Arrhenian behavior of the diffusion
coefficient,
\begin{equation}
\Gamma_0 \{ \exp (-\beta E_i)\}^{-1} \{ \exp(-\beta E_{ij})\}
  = \Gamma_0 \exp(-\beta E_{comb})                       \label{req}
\end{equation}
with a temperature-independent activation energy $E_{comb} \ge 0$.

Let us assume that $\nu_T(E)$ is restricted to the energy interval
[$-E_c,0$] with $E_c$ a positive quantity. If we identify $E_c$ with
$E_{comb}$ we can show by simple manipulations of the left-hand side
of Eq.\ (\ref{req}) that it is satisfied when
\begin{equation}
\nu_B(E) = \nu_T(E - E_{comb}).
\label{shift}
\end{equation}
The barrier distribution is then restricted to the interval
[$0,E_{comb}$] and it is simply the trap distribution shifted by
$E_{comb}$.
The argument implies that complete compensation, in $d \rightarrow
\infty$, is only possible for energy distributions that are restricted
to a finite interval. Of particular interest are distributions that
are symmetric about the midpoint of the interval. Of course it is not
necessary that the distributions are unequal zero on the whole
interval.

For finite dimensions $1 < d < \infty$ the diffusion coefficient
$D_{RB}$ has to be determined by approximations, for instance by the
EMA. It is a complicated functional of the energy distribution
$\nu_B(E)$ and no general result can be obtained. It seems very
implausible that complete compensation can be achieved in general.
However, it appears always possible, in dimensions $d \ge 3$,
to choose the energy distributions
$\nu_T(E)$ and $\nu_B(E)$ in such a way that an approximate Arrhenian
behavior is achieved in a restricted temperature interval.
\section{Comparison of EMA results and numerical simulations}
In this section we present EMA results for the combined model in $d
=3$ and 5 for different forms of the disorder and various temperatures.
The results are compared with Monte-Carlo simulations of the diffusion
coefficient by measuring the
mean-square displacement $\{ \vec{r}^{\,2}(t) \} $ where the asymptotic
diffusion coefficient $D$ is given by
\begin{equation}
\left\{ \vec{r}^{\,2}(t) \right\} \rightarrow 2dDt.
\end{equation}
Actually we present the whole proportionality factor $2dD$ as the
diffusion coefficient in the figures
to make results in different dimensions comparable.
Compensatory effects between random barriers and
random traps are now discussed.

We first calculate $D^{EMA}_{comb}$ for uniform distributions of
energies ( for the RB model see \cite{berna}),
\begin{equation}
   D^{EMA}_{comb} = \frac{\Gamma_0}{d-1} \exp(-\frac{1}{2}\beta
   \sigma_{B})
   \frac{\sinh(\frac{d-1}{2d}\beta \sigma_B)}{\sinh(\frac{1}{2d}
   \beta\sigma_B)}
   \frac{\beta \sigma_T}{\exp(\beta\sigma_T)-1},\quad d \geq 2
\label{gemauni}
\end{equation}
where
\[
  \nu(E) = \left\{
    \begin{array}{r@{\quad\quad}l}
       \frac{1}{\sigma_T} &  -\sigma_T \leq E \leq 0 \quad \mbox{(RT)}\\
       \frac{1}{\sigma_B} &  0 \leq E \leq \sigma_B \quad \mbox{(RB)} \\
       0 & \mbox{otherwise} 
    \end{array} \right. .
\]
By comparison with the numerical simulations we find that using
\begin{equation}
    \sigma_T \approx \frac{d-1}{d} \sigma_B
\label{gstsb}
\end{equation}
approximate Arrhenian behavior is reached.
In figure 2 results of Monte Carlo
simulations are shown together with the EMA  results (\ref{gemauni})
with parameters chosen
according to (\ref{gstsb}). The upward curvature for
the RB model is stronger in $d=5$ than in $d=3$ although the width of
the distribution is the same. The compensation works in the
case of a stronger curvature for the RB and RT models because
for larger coordination numbers compensation becomes generally easier.
The effect of the coordination number on the possibility of compensation
of barrier and site-energy disorder was already noted in \cite{LB}.

We consider as a second energy distribution the Gaussian distribution,
\begin{equation}
\nu (E) =
\frac{1}{0.95\sqrt{2\pi}\sigma}\exp\left(-\frac{(E-\bar{E})^2}
{2\sigma^2}\right)
\quad \mbox{with} \quad E \left\{
   \begin{array}{r}
     \le 0, \; \sigma = \sigma_T \quad \mbox{for RT} \\
     \ge 0, \; \sigma = \sigma_B \quad \mbox{for RB}
   \end{array} \right.
\end{equation}
We have to cut off
the tails of one side of the distributions to get only
negative respectively positive energy values. We decided to cut off
5\% of the distributions. Therefore we take $\bar{E}$
according to
\begin{equation}
\bar{E} = -\sigma \mbox{erf}^{-1}(0.95)
\end{equation}
where erf() is the error function.
For this distributions the self-consistency condition (\ref{self})
has to be solved numerically. Results are shown in figure 3.
For this example the two curves for the RT model lie on top of
each other because we took the same width in $d=3$ and $5$ and then
determined the width of the barrier distribution to get compensation
in the combined model. Again the upward curvature of the RB model
is stronger in $d=5$ than in $d=3$ although we took a smaller width of
the barrier distribution in the higher dimension. Nonetheless
compensation of
the curvature is still possible and the numerical procedure to find
the correct parameters of the distribution for the best possible
compensation is even more stable in the higher dimension. 
Though the curves of the combined
model in figure 3 look linear, the numerical results
show that no perfect compensation is possible.
\section{Conclusion}
Our theory of the diffusion coefficients of models with combined
site-energy and barrier disorder was based on weighted
transition rates where the thermal site occupancies enter explicitly.
Due to the exponential form of the assumed
Arrhenius law for the individual transition rates, the contributions
from the site energies and the barrier energies factorize.
If the site-energy and barrier disorder are uncorrelated, the resulting
diffusion coefficient factorizes into the
random-trap and random-barrier contributions. A consequence is that
compensation of the effects of random site-energies
and random barriers on the curvature in an Arrhenius plot of $D_{comb}$
versus
$\beta$ is not possible in $d = 1$ and $d = 2$. The simple-square
lattice represents a boundary case.

Partial compensation is possible in a finite temperature interval
in 3 and higher dimensions, if the strength of the
disordered site and barrier energies is
adjusted properly (stronger disorder in the barriers than in the
site-energies for finite dimensions $d \ge 3$).

In this context we should mention that our conclusions refer to the
hypercubic lattices (square, simple-cubic, etc.). The relevant
parameter for the temperature dependence of the diffusion coefficient
in the RB model is the coordination number $z$. Lattices with $z \ne 2d$
were investigated by Limoge and Bocquet \cite{LBtobep}.

We conclude that partial compensation
of the effects of random barriers and random site
energies is possible in $d = 3$, in a finite temperature interval,
if two premises are fulfilled,
\begin{description}
\item[i)] Assumption of independent site energies and barrier heights,
\item[ii)] Properly adjusted strength of the disorder.
\end{description}
Point i) may be somewhat alleviated by including short-range
correlations, but point ii) seems to be generally
necessary. We leave the question open whether the points i) and ii)  are
reasonable descriptions of real amorphous substances.

Instead we point to one serious deficiency of the present theories of
diffusion in disordered system. The present theories
are based on regular lattices with the disorder put into the transition
rates.
Real materials have topologically
disordered structures. There exist a range of coordination numbers of the
equilibrium sites, and the jump distances vary
considerably. The treatment of these effects remains as
a task for the future.

\acknowledgments
Discussions with J.L. Bocquet and Y. Limoge are gratefully acknowledged.

\newpage
\begin{figure}
\caption[]{Schematic representation of the combination of random
barriers and random traps.} 
\end{figure}
\begin{figure}
\caption[]{Diffusion coefficient in the RT, RB, and combined
model, for uniform distributions of the energies. 
The parameter $\Gamma_0$ is set unity. The different
symbols represent simulation results for the RB ($+$),
RT ($\diamond$), and the combined model ($\Box$)
with $\sigma_T = 3.0,3.2$ for $d = 3$ (full curves) and $d = 5$
(dashed curves) respectively and
$\sigma_B =4.0$ in both cases. The curves represent the EMA 
result for $2dD_{comb}$ (\ref{gemauni}).}
\end{figure}
\begin{figure}
\caption[]{Diffusion coefficient in the RT, RB, and combined
model, for the Gaussian distributions of energies. 
The parameter $\Gamma_0$ is set unity. The different
symbols represent simulation results for the RB ($+$),
RT ($\diamond$), and the combined model ($\Box$)
with $\sigma_B = 4.0$, $\bar{E} \approx 6.58$ for $d = 3$ (full
curves), $\sigma_B \approx 1.74$, $\bar{E} \approx 2.86$ for $d = 5$
(dashed curves), and
$\sigma_T = 1.2$, $\bar{E} \approx -1.97$ in both cases. The curves
represent the EMA result for $2dD_{comb}$ (\ref{gemauni}).}
\end{figure}
\clearpage
\renewcommand{\thefootnote}{\fnsymbol{footnote}}
\pagestyle{empty}
\begin{figure}[p]
\epsffile{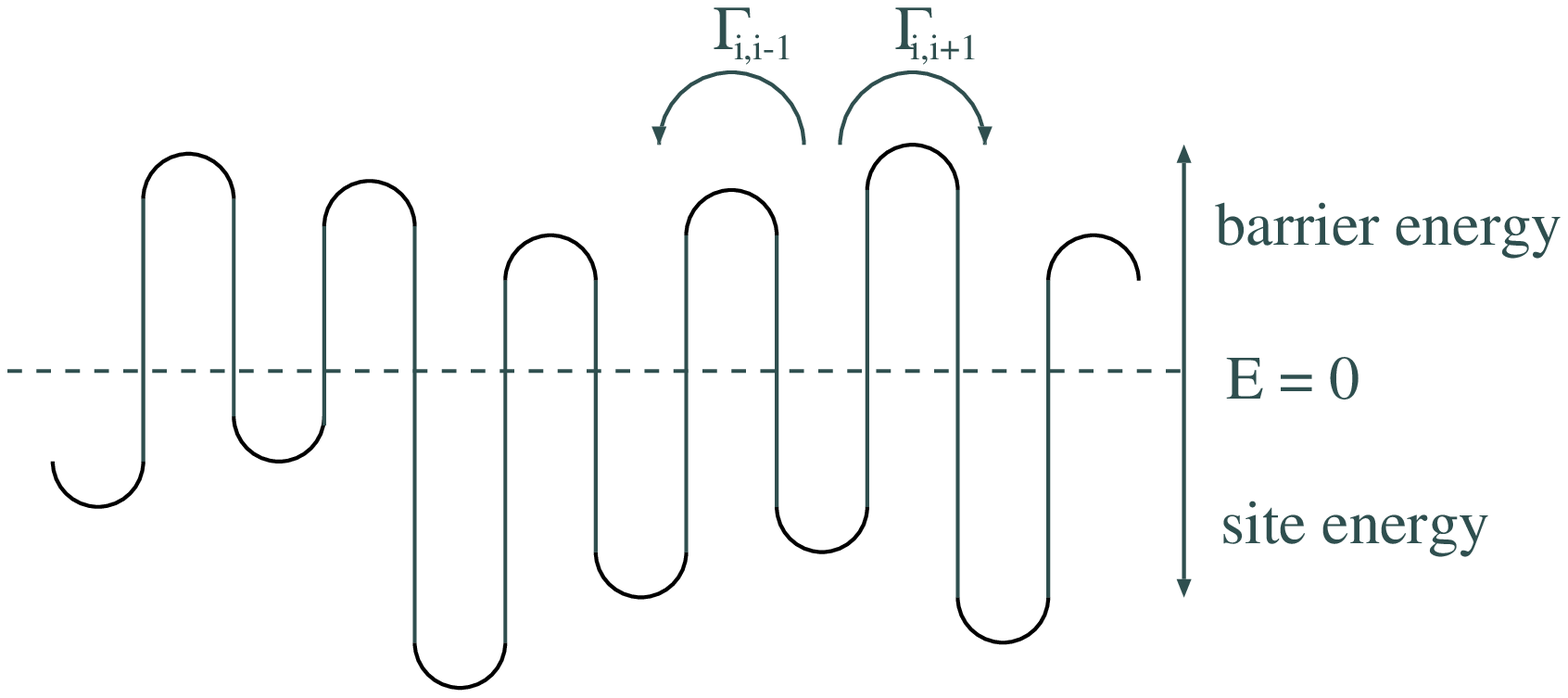}
\end{figure}
\footnotetext[2]{\normalsize Fig.\ 1}
\clearpage
\begin{figure}[p]
\epsffile{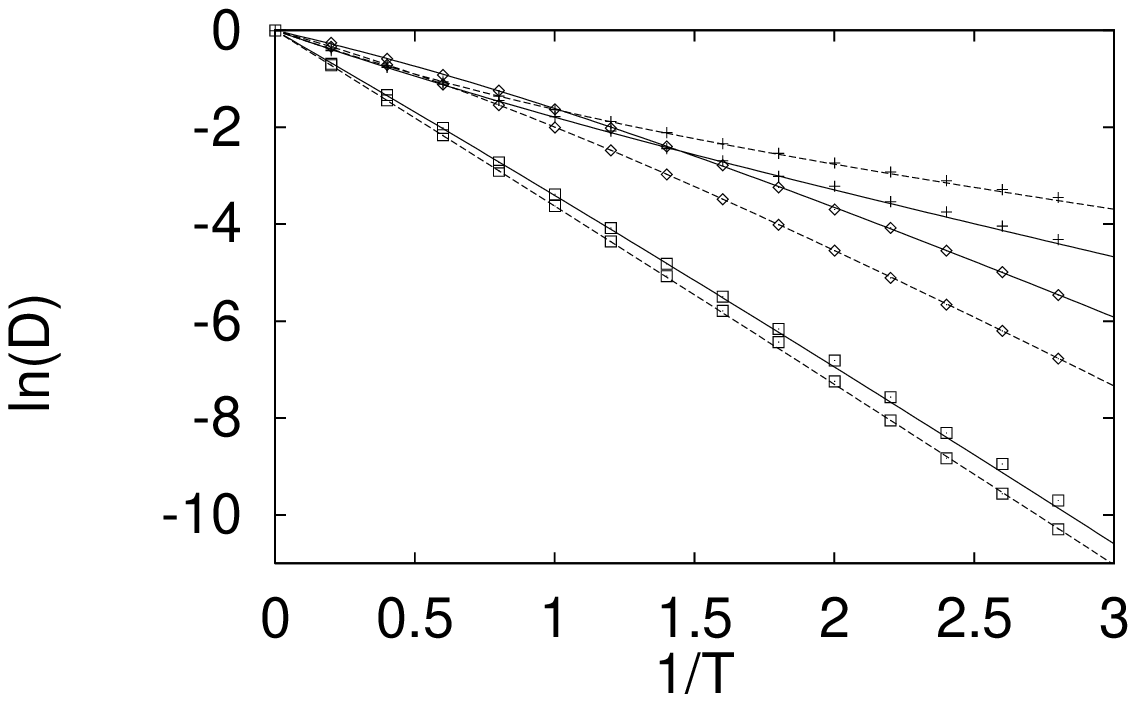}
\end{figure}
\footnotetext[2]{\normalsize Fig.\ 2}
\clearpage
\begin{figure}[p]
\epsffile{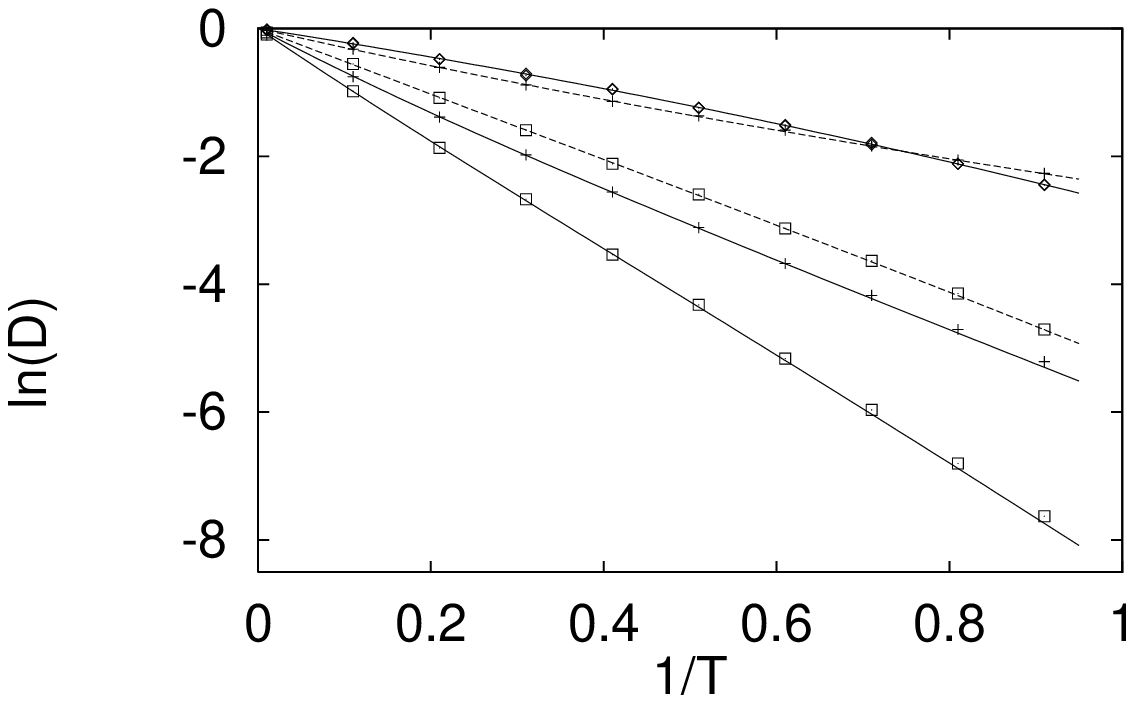}
\end{figure}
\footnotetext[2]{\normalsize Fig.\ 3}
\end{document}